\documentclass[fleqn,usenatbib]{mnras}
\usepackage{cancel,soul}
\usepackage{newtxtext,newtxmath}

\usepackage[T1]{fontenc}

\DeclareRobustCommand{\VAN}[3]{#2}
\let\VANthebibliography\thebibliography
\def\thebibliography{\DeclareRobustCommand{\VAN}[3]{##3}\VANthebibliography}

\usepackage{graphicx}	
\usepackage{amsmath}	
\usepackage{subfigure}
\usepackage{orcidlink}

\usepackage{ulem,xcolor}

\title[Detecting FRB by DANCE]{Detecting FRB by DANCE: a method based on DEnsity ANalysis and Cluster Extraction}

\author[Mao Yuan et al.]{
Mao Yuan\,\orcidlink{0000-0003-1874-0800}$^{1,2}$\thanks{E-mail: yuanmao@nssc.ac.cn},
Jiarui Niu\,\orcidlink{0000-0001-8065-4191}$^{3}$,
Yi Feng\,\orcidlink{0000-0002-0475-7479}$^{4}$,
Xu-ning Lv$^{1,2}$,
Chenchen Miao\,\orcidlink{0000-0002-9441-2190}$^{5}$,
Lingqi Meng\,\orcidlink{0000-0002-2885-568X}$^{3}$,\newauthor
Bo Peng\,\orcidlink{0000-0001-6956-6553}$^{3}$,
Li Deng$^{1,2}$,
Jingye Yan\,\orcidlink{0000-0002-7597-7663}$^{1,2}$\thanks{E-mail: yanjingye@nssc.ac.cn},
Weiwei Zhu\,\orcidlink{0000-0001-5105-4058}$^{3,6}$\thanks{E-mail: zhuww@nao.cas.cn}
\\
$^{1}$State Key Laboratory of Solar Activity and Space Weather, National Space Science Center, Chinese Academy of Sciences, Beijing 100190, China\\
$^{2}$Radio Science and Technology Center ($\pi$ Center), Chengdu 610041, China\\
$^{3}$National Astronomical Observatories, Chinese Academy of Sciences, Beijing 100101, China\\
$^{4}$Zhejiang Lab, Hangzhou, Zhejiang 311121, China\\
$^{5}$Qilu Normal University, College of Physics and Electronic Engineering, No. 2 Wenbo Road, Zhangqiu District, Jinan, China\\
$^{6}$Institute for Frontiers in Astronomy and Astrophysics, Beijing Normal University, Beijing 102206, China
}

\date{Accepted XXX. Received YYY; in original form ZZZ}

\pubyear{2025}

\begin{document}
\label{firstpage}
\pagerange{\pageref{firstpage}--\pageref{lastpage}}
\maketitle

\begin{abstract}
Fast radio bursts (FRBs) are transient signals {exhibiting diverse strengths and emission bandwidths}. Traditional single-pulse search techniques are widely employed for FRB detection; yet weak, narrow-band bursts often remain undetectable due to low signal-to-noise ratios (SNR) in integrated profiles. We developed DANCE, a detection tool based on cluster analysis of the original spectrum. It is specifically designed to detect and isolate weak, narrow-band FRBs, providing direct visual identification of their emission properties. This method performs density clustering on reconstructed, RFI-cleaned observational data, enabling the extraction of targeted clusters in time-frequency domain that correspond to the genuine FRB emission range. Our simulations show that DANCE successfully extracts all true signals with SNR~$>5$ and achieves a detection precision exceeding 93\%. Furthermore, through the practical detection of FRB 20201124A, DANCE has demonstrated a significant advantage in finding previously undetectable weak bursts, particularly those with distinct narrow-band features or occurring in proximity to stronger bursts.
\end{abstract}




\section{Introduction}
\label{sec:int}
Fast radio bursts (FRBs) have captivated the astronomical community due to the unresolved questions about their origins and the complexities of their emission mechanisms \citep{zhang2020physical,petroff2022fast,lyubarsky2021emission,zhang2023physics}. {Although} it is widely {accepted} that magnetars are the source of at least {a subset of} FRBs \citep{bochenek2020fast}, alternative models--including those invoking non-neutron star progenitors--have also been proposed (i.e. see the review in \citealt{zhang2023physics}). {Investigations into} energetics, and burst morphology, particularly for repeating sources, provide crucial insight into the underlying physical processes \citep{gourdji2019sample,beniamini2020periodicity,rajwade2020possible,chime2020periodic,cruces2021repeating,amiri2021first,li2021bimodal,pleunis2021fast,niu2022fast,zhang2022fast,zhou2022fast,zhu2023radio}. These studies benefit greatly from a larger sample of detected bursts; thus, detecting as many FRB bursts as possible is critical for {revealing} potential emission periodicities and constraining flux distributions.

{A variety of} search methods have been developed to detect both one-off and repeating FRBs. The most widely used pipelines are based on single-pulse search techniques \citep{cordes2003searches,spitler2014fast,champion2016five,zackay2017accurate,petroff2019fast,men2024transientx}. These methods involve de-dispersing data over various dispersion measures (DMs) and {evaluating} the signal-to-noise ratio (SNR) of integrated profiles for each DM trial.
Two main strategies are used to identify candidates from de-dispersed signals: (1) {selecting events} with sufficiently high SNR--usually above a threshold of 5 \citep{zhang2018fast}--and (2){locating} true pulses in DM-time diagrams. In the latter approach, a dispersed celestial burst produces prominent, centralized structures in a DM versus time scatter plot \citep{cordes2003searches} or a dedispersion transform space \citep{zackay2017accurate}.

The traditional SNR-based selection and DM search approaches, though widely adopted, often have limited detection sensitivity when {striving} to ensure true positive identification.  The identification of genuine signals within DM-transformed data is frequently impeded by impulsive radio frequency interference (RFI), particularly for pulses with low dispersion measures \citep{2009MNRAS.395..410E}. To mitigate this limitation, some methods split the data into multiple sub-bands to enhance narrowband FRB detection (e.g. \citealt{sand2022multiband,zhou2022fast}). 

Machine learning techniques offer a promising alternative for identifying features in FRB signals \citep{pang2018novel,zhang2018fast,men2019piggyback,aggarwal2021robust}.  
These techniques are generally applied either to DM-transformed data or directly to the original spectral, {exploiting} the unique characteristics of dispersed pulses. {In practical applications,} such methods have demonstrated higher sensitivity compared to conventional search methods \citep{zhang2018fast}. Nevertheless, despite these advances in automated detection, most pipelines--whether based on DM searches or feature-driven classification--still require visual inspection of pulse candidates. This secondary analysis in the time-frequency domain is essential for confirming low-confidence candidates and for conducting detailed investigations of the identified pulses. 

We {present} an unsupervised machine learning tool, DANCE, designed to identify both dispersed and de-dispersed FRBs in the spectrum. The method exploits temporal and spectral information without requiring prior feature training. This method utilizes density analysis and clustering to generate visually interpretable candidates, directly marking their time-frequency locations in the spectrum. For clustering, we use DBSCAN (Density-Based Spatial Clustering of Applications, \citealt{ester1996density}) to {distinguish} astrophysical signals from background noise. Similar techniques have been used in FRB detection \citep{pang2018novel,men2024transientx}, {where clusters are derived} from secondary derived features. Because cluster methods are particularly effective for detecting discrete point samples in diagram spaces, some studies have employed clustering to assemble single pulse events within DM-transformed space at optimal DMs \citep{foster2018verifying}. {In contrast, DANCE operates directly on the original spectral data,} identifying clusters in the time-frequency spectrum rather than in a DM transformation diagram.

The structure of the paper is organized as follows. In Sec.\ref{sec:method}, we introduce the methods, including descriptions of RFI mitigation, density analysis, and cluster extraction. Sec.\ref{sec:res} presents the performance of DANCE on both simulated FRB signals and real observations of FRB 20201124A. In Sec.\ref{sec:dis} and Sec.\ref{sec:con}, we discuss the extended application of DANCE and make a conclusion of our work. 

\section{Methods}
\label{sec:method}
Our method {exploits} the fact that celestial signals, like FRBs, occupy {localized regions in the time–frequency plane and exhibit power levels significantly higher} than their surrounding noises. By filtering out lower-power samples, the target signal region becomes denser than the surrounding filtered data. Subsequent density analysis \st{methods} then enable us to distinguish the signal location from background noise within the spectrum.

The primary aim of DANCE is to project the time-frequency representation into a signal density space (SDS) and subsequently identify regions of higher density within this space. The SDS is represented as a binary matrix with dimensions identical to those of the original time–frequency representation. For each pixel $p_{ij} = (t_{i}, f_{j}$) in the SDS, $p_{ij} =1$ if {the corresponding pixel in the spectrum exceeds a defined power threshold}; otherwise  $p_{ij} = 0$. The distribution of positive pixels thus characterizes the signal density derived from the spectrum, where genuine celestial signals, such as FRBs, are expected to produce a substantially higher concentration of positive pixels than random noise. The integrated flux in the signal's frequency band consequently exhibits a distinct peak. DANCE is specifically designed to detect these high-density regions and to extract potential FRB candidates.

\begin{figure}
\includegraphics[width=\columnwidth]{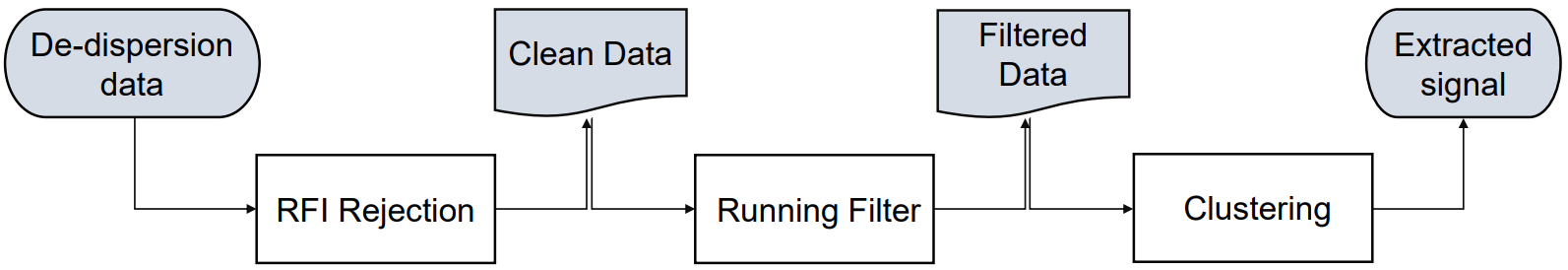}
\caption{The detection process.}
\label{fig1}
\end{figure}

As illustrated in Fig.\ref{fig1}, the detection process comprises three principal stages: (1) RFI mitigation using a two-dimensional discrete wavelet transform (2D-DWT), (2) signal filtering via a running-mean algorithm, and (3) signal extraction through DBSCAN clustering.

\subsection{Two-dimensional discrete wavelet transform}
{We employ the 2-D DWT to obtain RFI-mitigated data and identify contaminated regions in the time-frequency spectrum.} The one-dimensional discrete wavelet transform (1-D DWT) decomposes a discrete signal series,$ S[n] $, where $ n = 0, 1, \dots, N-1 $, into approximation and detail at multi-resolution. {The 2-D DWT extends these principles to a two-dimensional time-frequency representation, $ S[m, n] $, enabling independent decomposition along the time ($ m $) and frequency ($ n $) axes.}

{For the 1-D DWT, a discrete signal, $ S[n] $, is decomposed using a scaling function $ \phi(t) $ and a wavelet function $ \psi(t) $. The approximation and detail coefficients at resolution level $ j_0 $ and higher are defined as:
\begin{align}
\begin{aligned} 
l(j_0, m) &= \sum_{n=0}^{N-1} S[n] \phi_{j_0, m}(n), \\
h(j, m) &= \sum_{n=0}^{N-1} S[n] \psi_{j, m}(n) \quad (j \geq j_0),
\end{aligned}
\end{align}
where $ \phi_{j_0, m}(n) = 2^{j_0/2} \phi(2^{j_0} n - m) $ and $ \psi_{j, m}(n) = 2^{j/2} \psi(2^j n - m) $ represent the scaled and translated scaling and wavelet functions, respectively. The scaling function acts as a low-pass filter, producing approximation coefficients $ l(j_0, m) $ that capture the coarse signal structure, while the wavelet function acts as a high-pass filter, yielding detail coefficients $ h(j, m) $ that represent high-frequency components \citep{mallat1999wavelet}. In practice, the 1-D DWT is implemented using a filter bank with low-pass and high-pass filters followed by downsampling, as described by the fast wavelet transform \citep{strang1996wavelets}.}

{The 2-D DWT extends the same framework to a time-frequency representation $S[m, n] $, where $ m = 0, 1, \dots, M-1 $ and $ n = 0, 1, \dots, N-1 $ denote time and frequency indices, respectively. }
The 2-D  scaling and wavelet functions, $\phi (m,n)$ and $\psi (m,n)$, follow the same propitiates as the 1-D cases. In practice, users commonly choose separable functions to decompose the time-frequency representation (with time and frequency samplings $m$ and $n$). Following this idea, the 2-D form contains one scaling function and three wavelet functions, they are
\begin{align}
\begin{aligned} 
    \phi^{ll}_{j0} ({m,n}) & = \phi_{j0}({m}) \phi_{j0}({n}),   \\
    \psi^{hl}_{j}({m,n}) & = \psi_{j}({m}) \phi_{j}({n}), \\
    \psi^{lh}_{j}({m,n}) & = \phi_{j}({m}) \psi_{j}({n}), \\
    \psi^{hh}_{j}({m,n}) & = \psi_{j}({m}) \psi_{j}({n}). 
\end{aligned}
\end{align}
Then we get four groups of coefficients through 2-D DWT, they are 
\begin{align}
\begin{aligned} 
    LL (j0,m,n) & =\sum_{{m}=0}^{M-1}\sum_{{n}=0}^{N-1} S({m,n}) \phi^{ll}_{j0} ({m,n}),  \\
    HL (j0,m,n) & = \sum_{{m}=0}^{M-1}\sum_{{n}=0}^{N-1} S({m,n}) \psi^{hl}_{j}({m,n}),  \\
    LH (j0,m,n) & = \sum_{{m}=0}^{M-1}\sum_{{n}=0}^{N-1} S({m,n}) \psi^{lh}_{j}({m,n}),  \\
    HH (j0,m,n) & = \sum_{{m}=0}^{M-1}\sum_{{n}=0}^{N-1} S({m,n}) \psi^{hh}_{j}({m,n}).  
\end{aligned}
\end{align}

Among these four groups of coefficients, {The $ LL $ coefficients represent the approximation of $ S[m, n] $, while the $ HL $, $ LH $, and $ HH $ coefficients capture horizontal, vertical, and diagonal details, respectively. The 2-D DWT is computed by applying 1-D DWT sequentially along the time and frequency axes, involving filtering and downsampling operations \citep{strang1996wavelets}.}

{In our application, a multi-resolution 2-D DWT decomposes the time-frequency representation into subbands at different scales, exposing RFI as localized high-amplitude coefficients in the detail subbands ($ HL $, $ LH $, and $ HH $). RFI-mitigated data are obtained by applying thresholding techniques (e.g., soft or hard thresholding) to the wavelet coefficients and performing the inverse 2-D DWT \citep{offringa2010post}. The choice of wavelet basis (e.g., Daubechies or Haar) and decomposition levels depends on the RFI characteristics and is typically determined empirically.} 

\subsection{Density analysis}
We employ a density analysis to identify FRB signals directly within the time-frequency spectrum, {rather than} relying on the SNR of the integrated profile. This approach consists of two key procedures: signal filtering and density clustering.

Signal filtering is designed to identify ``positive'' signals in the time-frequency space. In this context, a ``positive'' signal refers to a sample whose power exceeds a predetermined threshold in the spectrum. Specifically, if the power of a sample surpasses this threshold, its corresponding time-frequency location is considered a true value; otherwise, it is classified as a false value. This method effectively transforms the power spectrum into a binary dataset, wherein the density of true signal positions in the vicinity of an FRB is expected to be significantly greater than in other areas. Nevertheless, if the FRB signal is weak (i.e., with a low SNR in the integrated profile), both desired signals and noise (or omitted RFI) can be marked as positive signals. We perform a second filter to the RFI-filtered data, the method we use is the running mean. The running mean operation averages the energy of isolated outliers within a limited window, helping to reduce noise and smooth outliers by sliding the window along the data series. Mathematically, the smoothed signal is 
\begin{equation} 
\centering
s_{\mathrm{new}}(i) = \{\frac{1}{k} \sum_{j=0} ^{j=k-1}s (i+j),\,i=0,1,2,\cdots,n-k-1\},
\end{equation} 
Given that DANCE focuses on extracting only the spectra area of an FRB within the signal density space (SDS), we apply the running mean window exclusively within the bandpass for each temporal recording. We then filter out signals that fall below a specified threshold and flag the positions of positive signals in the time-frequency space. The result of this filtering is a binary dataset, where positive signal locations are represented by 1 and false cases by 0, visually depicted as points in the time-frequency space.

After obtaining the density distribution of positive signals in the time-frequency space, our subsequent objective is to cluster the dense zones. We employ the DBSCAN method to identify potential dense clusters. DBSCAN is a density-based technique to discover clusters in a dataset with both designed signals and noise \citep{ester1996density,khan2014dbscan}, it is a supervised machine learning method without designing the cluster numbers and types. The leading theories of DBSCAN are density-reachability and density-connectivity \citep{ester1996density}. Two key parameters dominate the clusters DBSCAN can find, the neighborhood radius of a point \textit{Eps} and the minimum samples in the neighborhood radius, \textit{MinPts}.
For any sample $p$ in a DBSCAN-identified cluster $C$, it must satisfy two conditions,
\begin{align}
\begin{aligned}
    Num_{\mathrm{Eps}}^{\mathrm{p}} &>= MinPts, \\
    dist_{\mathrm{p,q}} &<= Eps.
\end{aligned}
\end{align} 
The first relation indicates a density-reachable condition, while the second relation pertains to density-connected conditions. Among which, $Num_{\mathrm{Eps}}^{\mathrm{p}}$ is the sample number of a circle centered at $p$ with a radius is \textit{Eps}, $dist_{{p,q}}$ is the distance between $p$ and its any neighboring samples $q$.
FRB-signal regions in the smoothed time-frequency space are dense and continuous in comparison with the noise.  Therefore a density-reachable and density-connected clustering can find them well. In addition to the designed signals, noise can also be grouped into various discrete and small-size clusters. Setting large \textit{Eps} and $dist_{p,q}$ prevents clustering the sparse noise, but that may also leave out some weak FRBs. We choose small \textit{Eps} and $dist_{p,q}$ for DANCE.

\subsection{Cluster extraction}
Dense analysis can generate various clusters when we choose an easy density-connected condition. We need to gather the designated cluster samples and accurately identify the actual location of the FRB using the appropriate clusters. Cluster prominence is a valid criterion since only astronomical signals that have a significant range of frequencies and a wide pulse duration are capable of forming extensive and compact clusters. We can calculate the pixel number of a cluster sample to measure its prominence and extract the samples with substantial coverage in the SDS by applying a statistical criterion to all cluster sizes. 

The extraction strategy is a Z-score transformation of the size statistic. We calculate the Z-score of all identified clusters and then filter the trivial samples. Mathematically, $Z$ can be calculated like
\begin{align}
Z = \frac{x-\mu}{\sigma},
\end{align}
where $x$ is the size statistic of all clusters, $\mu$ and $\sigma$ is the mean and standard deviation of $x$.
Prominent samples get high $Z$ and it is easy to extract them by a simple threshold filtering.

We also designed a double criterion to filter possible RFI or noise clusters. Because the SDS is RFI-clean and noise-smooth, omitted interference could be a baseline fluctuation or background noise. Hence signal pixels of them in SDS are widely distributed. An FRB pulse differs mainly in limited pulse width, as does its cluster. Therefore, we also calculate the characteristic width of previously extracted samples and exclude wide cases. The characteristic width is measured by the maximum range of the sample in the temporal dimension. 

\section{Processing Flow}
This section describes the data processing and burst signal extraction steps used by DANCE. We show the visual results of the key flows using the actual observation data from FAST. 
\subsection{RFI mitigation and denoising}
{The strategy of most RFI mitigation techniques} is flagging the RFI blocks and then zapping them \citep{agarwal2020initial,rafiei2023mitigating}. Such an operation helps to enhance the SNR of detection results in dedispersed integrated signals. However, a zapped spectrum causes substantial signal density shifts that may bias cluster formation. As we mentioned in Sec.\ref{sec:method}, DBSCAN is a density-based connectivity approach that ceases clustering when encountering gaps in the data, potentially preventing the formation of a continuous dense cluster for the FRB signal.

\begin{figure*}
\subfigure[raw data]{
\includegraphics[width=0.5\columnwidth]{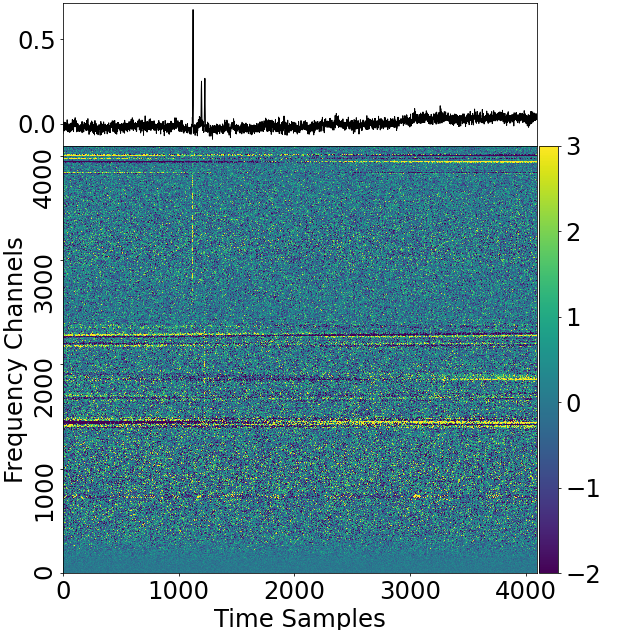}}
\subfigure[decomposed data]{
\includegraphics[width=0.5\columnwidth]{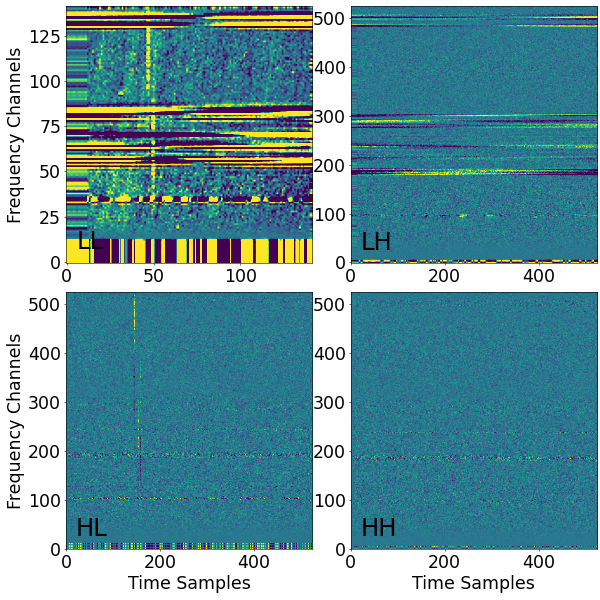}}
\subfigure[RFI flagging]{
\includegraphics[width=0.5\columnwidth]{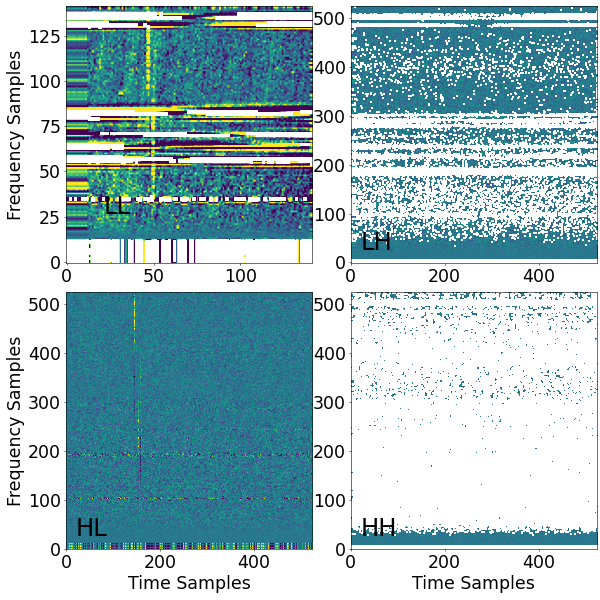}}
\subfigure[reconstructed data]{
\includegraphics[width=0.5\columnwidth]{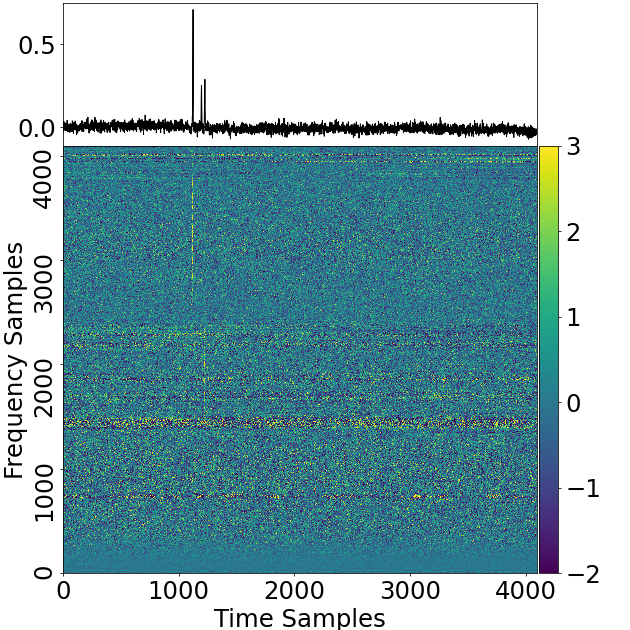}}
\caption{RFI mitigation using the 2-D DWT for strong FRB signals.  (a) The original data containing three FRB bursts, with the top panel showing the integrated signal. (b) The wavelet decomposition at the third level, displaying the four subcomponents ``LL, LH, HL, and HH'', respectively.  Among these, ``LL'' represents the low-frequency approximation of the raw data, ``HL, LH'' extract horizontal and vertical features of the data, respectively, and ``HH'' captures fine-scale details. The decomposition is performed over five levels, with the third-level components shown here as representative examples. (c) RFI identification, where blanked regions indicate flagged interference at a 1~$\sigma$ threshold. ``HL'' component is retained unflagged, as the desired FRB signals are primarily represented in this channel. (d)The reconstructed data obtained after RFI removal from the decomposed components. All panels share an identical color scale for direct comparison.}
\label{fig_rfi}
\end{figure*}

To demonstrate the use of 2-D DWT for RFI reduction, we use a dataset that exhibits significant interference. The frequency range covered by this data, which was observed by FAST, is 1000–1500~MHz, and the duration time is 6.24\,s. 
Both frequency and time samples in the dataset are 4096. We removed the baseline in the time dimension by deducting the mean power of each channel.
Fig.\ref{fig_rfi} shows the RFI mitigation procedures. We first decompose the data into multi-levels. In practice, the pixel resolution of the decomposed picture decreases in the form of $2^{l}$ where $l$ is the level. For example, the original data size is 4096$\times$4096 in time and frequency samples. When the decomposition level is $l=5$, the size of the ``LL'' component should be 128$\times$128, while the other components (``LH'', ``HL'', and ``HH'') are $4096/2^{i} \times 4096/2^{i}$ which $i$ is the output layer, defines from 1 to 5. We then set a threshold filtering to all the layers, in which the values beyond $\pm 1 \sigma$ are replaced by 0. The blank data in panel (c) are the flagged outliers. Finally, we make a 2-D inverse DWT by using the filtered decomposition layers and derived reconstructed data, shown in panel (d). 

The 2-D DWT method meets our density analysis requirements and offers several benefits. First, it allows RFI elimination without deleting data blocks. Second, only filtering a subset of the deconstructed components keeps the desired signals. Even if RFI signals are weaker than FRB signals, we can indicate them without erasing the planned signals. The bias only occurs with clean data or strong FRB signals. 

\subsection{Density analysis}
\begin{figure}
\includegraphics[width=\columnwidth]{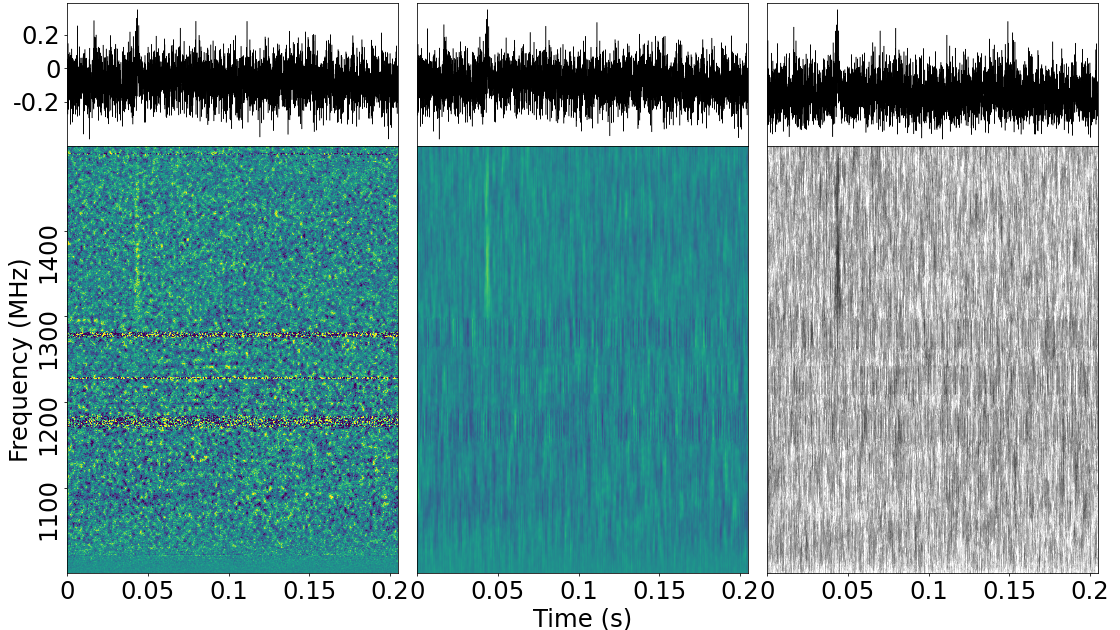}
\caption{Transformation from the reconstructed spectrum to the signal density space (SDS). The left panel shows the reconstructed data; the middle panel displays the smoothed data; and the right panel presents the corresponding signal density representation. A weak FRB signal is {visible within the region} at 0.04-0.045\,s in time and approximately 1300-1500\,MHz in frequency.}
\label{fig_density}
\end{figure}

The reconstructed spectrum is RFI-poor but is still not smooth due to some isolated outliers and noises, shown in Fig.\ref{fig_rfi} and Fig.\ref{fig_density} (left panel). As we introduced in Sec.\ref{sec:method}, outliers in the spectrum can contaminate the SDS.

We use a running filter to smooth the data and mitigate the fluctuations of noise. The middle panel in Fig.\ref{fig_density} is the smoothed signal of the original spectrum (left panel), which shows that the new spectrum is smooth except for the high-density zone which indicates an FRB signal. The width of the running window in this case is 256 frequency channels ($\sim$ 30~MHz). It is important to note that the burst bandwidth could be broadened after a smooth operation. The extension effect is dependent on both the width of the running window and the signal strength. We have a discussion about it in the next section.

The SDS is a presentation of pixel location where the signal power is over an artificial strength threshold. To evaluate the density of SDS, we apply a filtering process based on the Interquartile Range (IQR), a measure of statistical dispersion. The IQR is the range between the first quartile (25th percentile) and the third quartile (75th percentile) of a data set, effectively capturing the spread of the central 50\% of the data. The right panel in Fig.\ref{fig_density} is an SDS with a threshold at 1.5~IQR (beyond the 75th percentile of the signal power statistics). We perform a DBSCAN clustering analysis of the density distribution in SDS (Fig.\ref{fig_density}, right panel). As we introduced in Sec.\ref{sec:method}, \textit{Eps} and \textit{MinPts} determine how the DBSCAN defines a density-connected set. The \textit{MinPts} is an easier-set parameter for DBSCAN. A relatively small \textit{MinPts} helps include all possible clusters. Usually, setting \textit{MinPts} to twice the dimensionality of the data, i.e., $MinPts=4$, or near this value, is suitable for most density analysis cases \citep{sander1998density,schubert2017dbscan}. The other parameter \textit{Eps} is harder to set as it depends on the distance of the neighboring samples in the wanted clusters, which is a posterior measurement. Some researchers suggested adaptive mathematical approaches to calculate and get appropriate \textit{Eps} before performing clustering \citep{ester1996density,starczewski2020new}. 
For FRB clustering, we determine the \textit{Eps} and  \textit{MinPts} according to the FRBs' morphological characters. Being different from traditional point sets, the SDS are projection of a filtered spectrum. The majority of the points are locations of RFIs or noises. Clusters defined by them are random, sporadic, and small in size. On the contrary, the FRB clusters should be larger denser zones gathered with continuous points. Therefore we choose a larger \textit{MinPts}, thus helping to pass most noise clusters. We define a {dimensionless} density parameter $\rho$ to replace \textit{MinPts}, i.e. $MinPts = \lceil S \rho N_{0}\rceil$, in which $S$ (in pixel$^{2}$) is a clustering cell area determined by radius \textit{Eps}, and $\rho$ as the multiple of the average number of elements per pixel ($N_{0}$) in the SDS.

Considering that the narrow burst can exhibit a short duration within a timescale at tens of microseconds (or even narrower) \citep{farah2018frb,nimmo2020microsecond}, we set $Eps=2$ which is twice the sampling timescale (49.152\,$\mu s$) of the FAST. {By filtering the spectrum using 1.5 times the IQR, we remove outliers and focus on the core clusters in SDS. This results in a background point density of 0.25 per pixel.}Thus, the appropriate density parameter for DBSCAN is $1 \leq \rho \leq 4$. 

\begin{figure}
\includegraphics[width=\columnwidth]{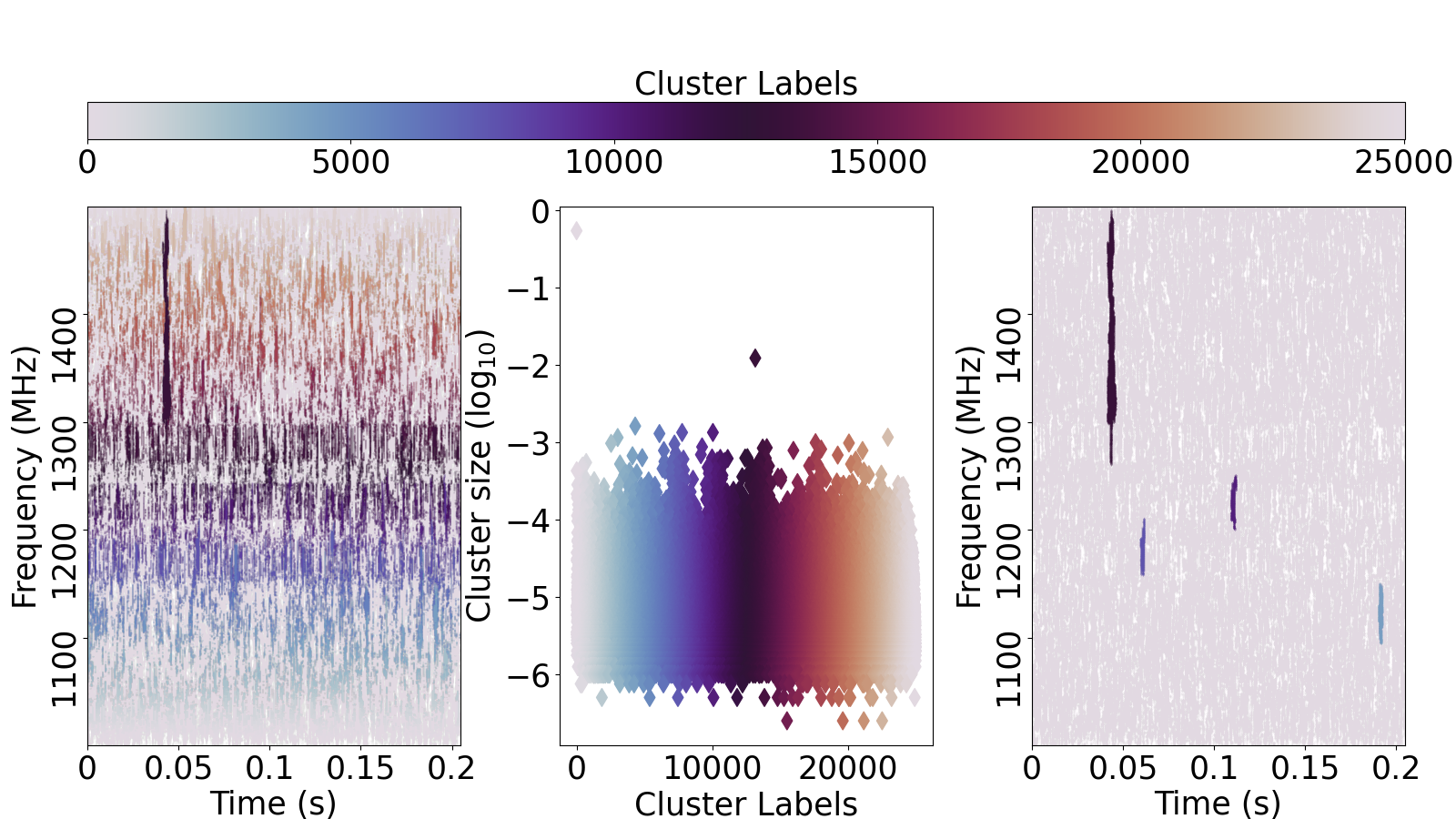}
\caption{DBSCAN-identified clusters. The top color bar indicates different clusters. Left: all separated cluster samples in the SDS of Fig.\ref{fig_density}, with each color representing a distinct cluster. Middle: the size of each cluster sample, measured by the proportion of pixels each cluster occupies. Two notable clusters are labeled -1 and 13180 in the color bar, with sizes of 54\% ($\log_{10} = -0.64$) and 1\% ($\log_{10} = -2$), respectively.  Right: the five largest cluster samples, including the background noise (label -1), an FRB signal (label 13180), and three other noise clusters.}
\label{fig_cluster}
\end{figure}

A larger $\rho$ helps to extract stronger signals, we suggested $\rho \sim2$ is suitable for clustering signals from noises as well as possible. 
Fig.\ref{fig_cluster} shows the result of FRB detection with DANCE. The left panel is the DBSCAN-identified clusters. Here the radius of the scan circle \textit{Eps} is set as 2 pixels. The density-reachable threshold, \textit{MinPts} is set as 2 times larger than the sample density of SDS. We totally identified 25,023 clusters in this SDS, and we get two prominent samples, which are labeled as -1 and 13,180, shown in the middle panel. We use the number of samples a cluster contains to indicate its size. Among them, the cluster with the label 13,180 is the projection of an FRB while the most prominent cluster is the global background noise.

\subsection{Cluster extraction}
\begin{figure}
\includegraphics[width=\columnwidth]{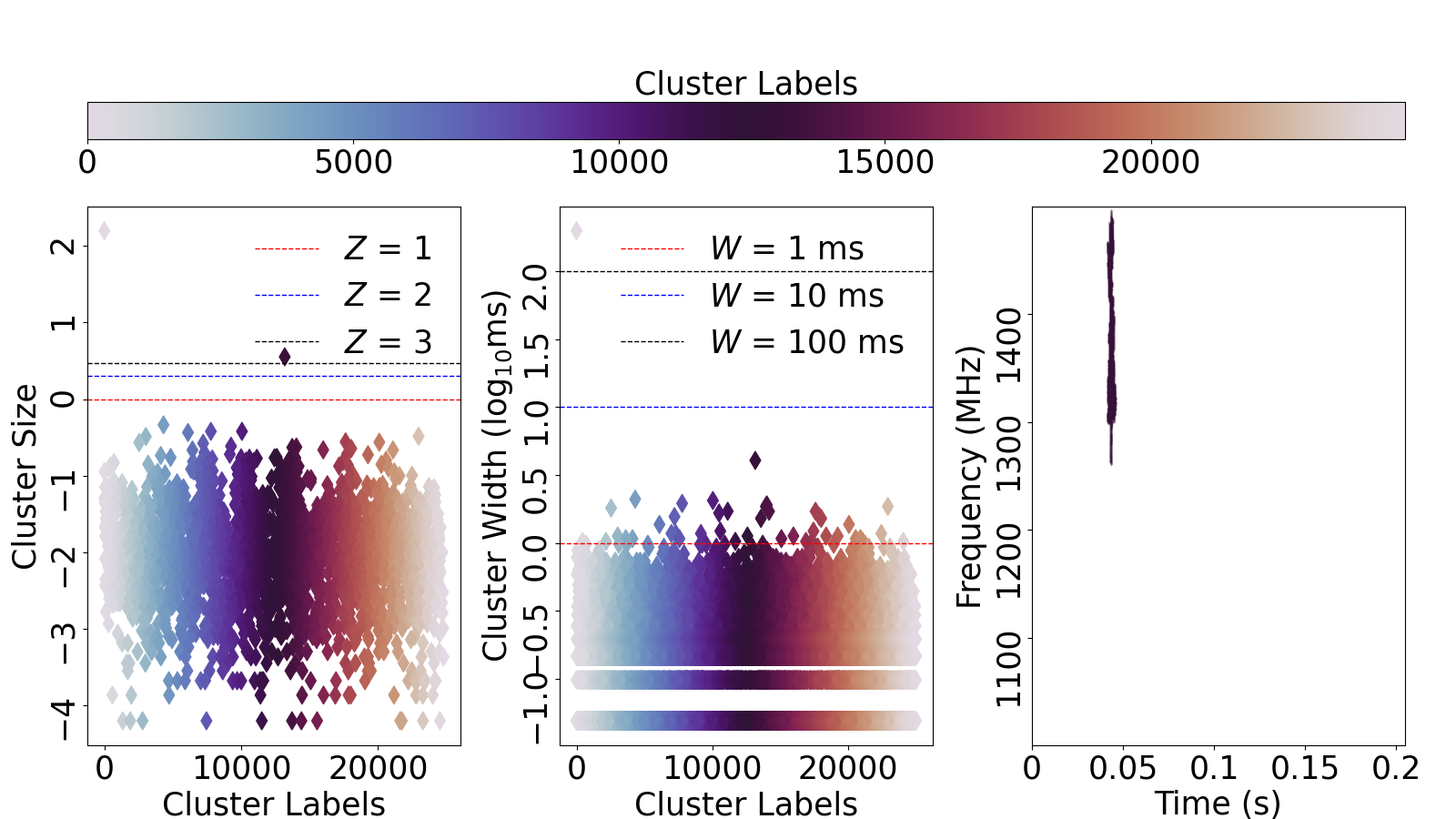}
\caption{Cluster extraction based on Z-score and pulse width. The color bars at the top indicate the cluster labels. Left: Z-score of all cluster sizes in the SDS of Fig.\ref{fig_density}, each color indicates a cluster. Middle: the pulse width of each cluster sample. Right: the filtered FRB cluster with a Z-score of 3.6 and a pulse width of 4.1~ms, respectively. This cluster has been filtered out based on a criteria of $Z>3$ and $W<100$.
}
\label{fig_cluster_t}
\end{figure}

We extract prominent clusters based on their sizes. A statistic of the cluster size is represented in Fig.\ref{fig_cluster}, the middle panel. We calculated the statistical Z-score ($Z$) of these cluster sizes, and $Z$ scores are shown in Fig.\ref{fig_cluster_t}, left panel. This case shows two prominent clusters with $Z>3$. We then calculate the characteristic pulse width of each cluster, shown in the middle panel of Fig.\ref{fig_cluster_t}. As we have introduced in {Sec.\ref{sec:method}}, secondary filtering is necessary to {remove} some RFI- or background noise-clusters. We use the scale of temporal coverage of a cluster to define its pulse width. The middle panels indicate that the width of the background noise-cluster covers the whole observation subband ($\sim$ 200~ms). 

Though vast sporadic noises have comparable widths with FRB signals (e.g. 1-10~ms), their sizes are trivial. We extract the FRB cluster by setting thresholds to cluster sizes and widths. In this case, we choose cluster samples with $Z>3$ and $W<100$~(ms). FRB signals are the only sample that has both a prominent size and reasonable pulse width, as shown in the fight panel. By using DANCE we can extract the weak FRB with more convincing visible detection results than the common integrated profile method.

\section{Results}
\label{sec:res}

\subsection{Detection on simulated FRBs}
We evaluate the capability of DANCE by detecting simulated FRBs in random noise. The experiment strategy is to produce stimulative FRB signals with various strengths in background data, and then use DANCE to extract the targets. By counting false and missing identification cases, we finally approximate the detection limits of DANCE. 

For the background data,  we generate a matrix with a dimension is 4096$\times$4096. The element values of the data block follow the Gaussian distribution with a mean value of 0 and a standard deviation of 1. We take the noise data as a spectrum background, in which the integration time and frequency bands are assumed at 200~ms and 500~MHz. For the simulated burst, we use a two-dimensional Gaussian function to generate FRB signals. Assuming the signal strength is $F(f,t)$ at $f$~MHz and $t$~ms, it has the form like 
\begin{align}
\begin{aligned}
F(f,t) = Ae^{-({\frac{(C_\mathrm{t}-t)^2}{2\sigma^2_\mathrm{t}}}+{\frac{(C_\mathrm{f}-f)^2}{2\sigma^2_\mathrm{f}}})}.
\end{aligned}
\end{align}
Here $A$ dominates the strength, i.e. the SNR, of the integrated pulse. $C_t$ and $ C_f$ indicate the center location of the signal at time and frequency, respectively. During the simulation, our progress randomly generated three parameters $A$, $\sigma_t$, and $\sigma_f$, which control the pulse strength, width, and bandwidth, respectively. 

\begin{figure}
\includegraphics[width=\columnwidth]{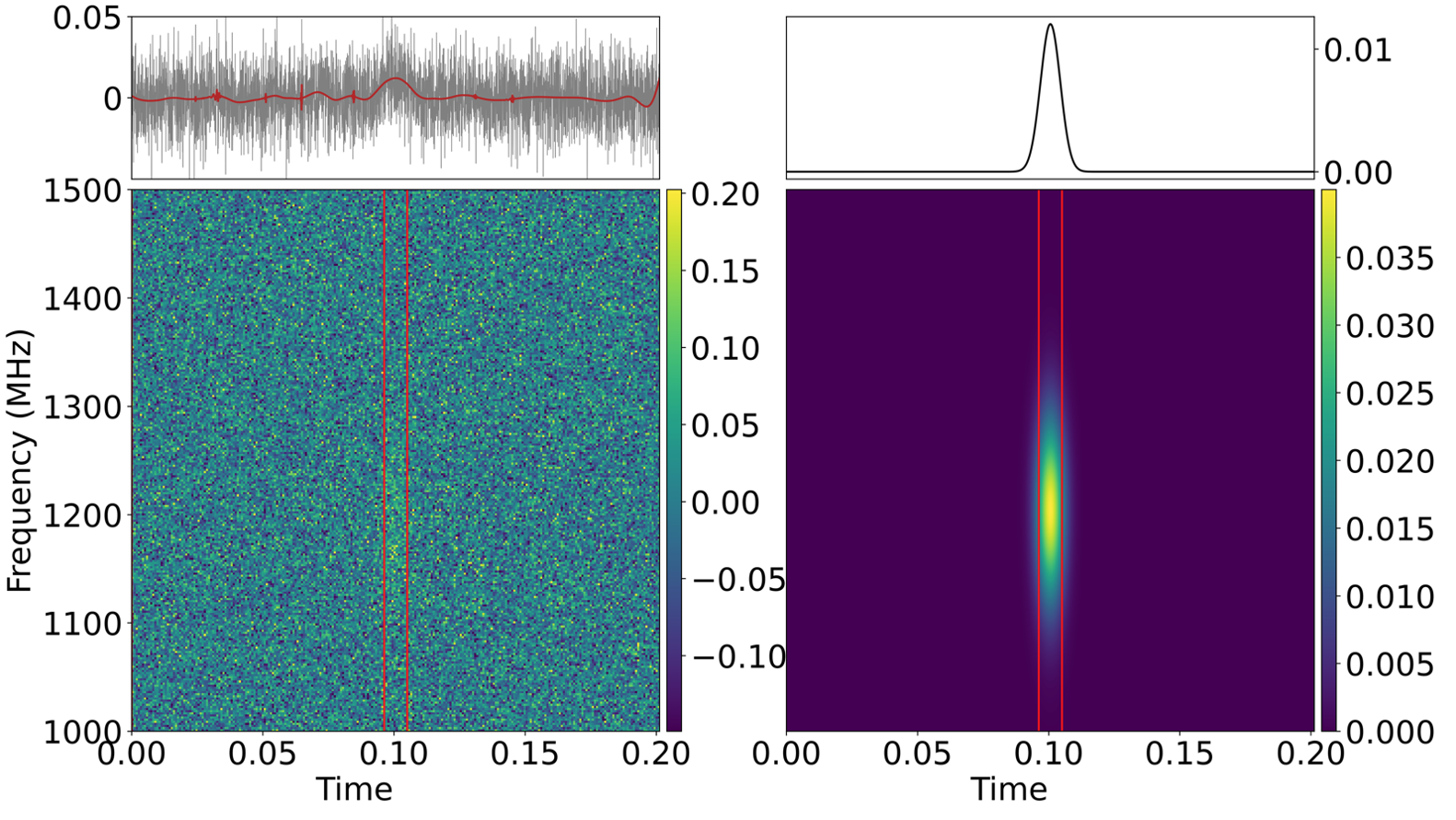}
\caption{A presentation of a simulated FRB. The left panel is the data mixed with noise and analog FRB signal. The right panel is the simulated FRB signal. The pulse width is 9~ms, which is denoted by the vertical red lines. The bandwidth is 140~MHz and the SNR of the combined signal is $\sim 8.3$.  }
\label{simulate}
\end{figure}

Fig.\ref{simulate} presents combined data (left panel) and the simulation data (right panel).  The artificial FRB in this case located at the center of the spectrum, with a pulse width of 9~ms and a bandwidth of 140~MHz. Here we define the pulse width $W_{\mathrm{p}}$ (in bins) as the full-width half maximum (FWHM) of the peak of the integrated simulated pulse. The SNR of the combined integrated signal is $\sim8.3$, which was calculated by
\begin{align}
SNR = \frac{1}{\sqrt{W_{\mathrm{p}}}\sigma}\sum_{i=1}^{n_{\mathrm{bins}}}p_{\mathrm{simu}},
\end{align}
where $\sigma$ is the standard deviation of the integrated time series of the background data, and $\sum_{i=1}^{n_{\mathrm{bins}}}p_{\mathrm{simu}}$ denotes the summed power of the simulated gaussian pulse within the FWHM.

\begin{figure}
\includegraphics[width=\columnwidth]{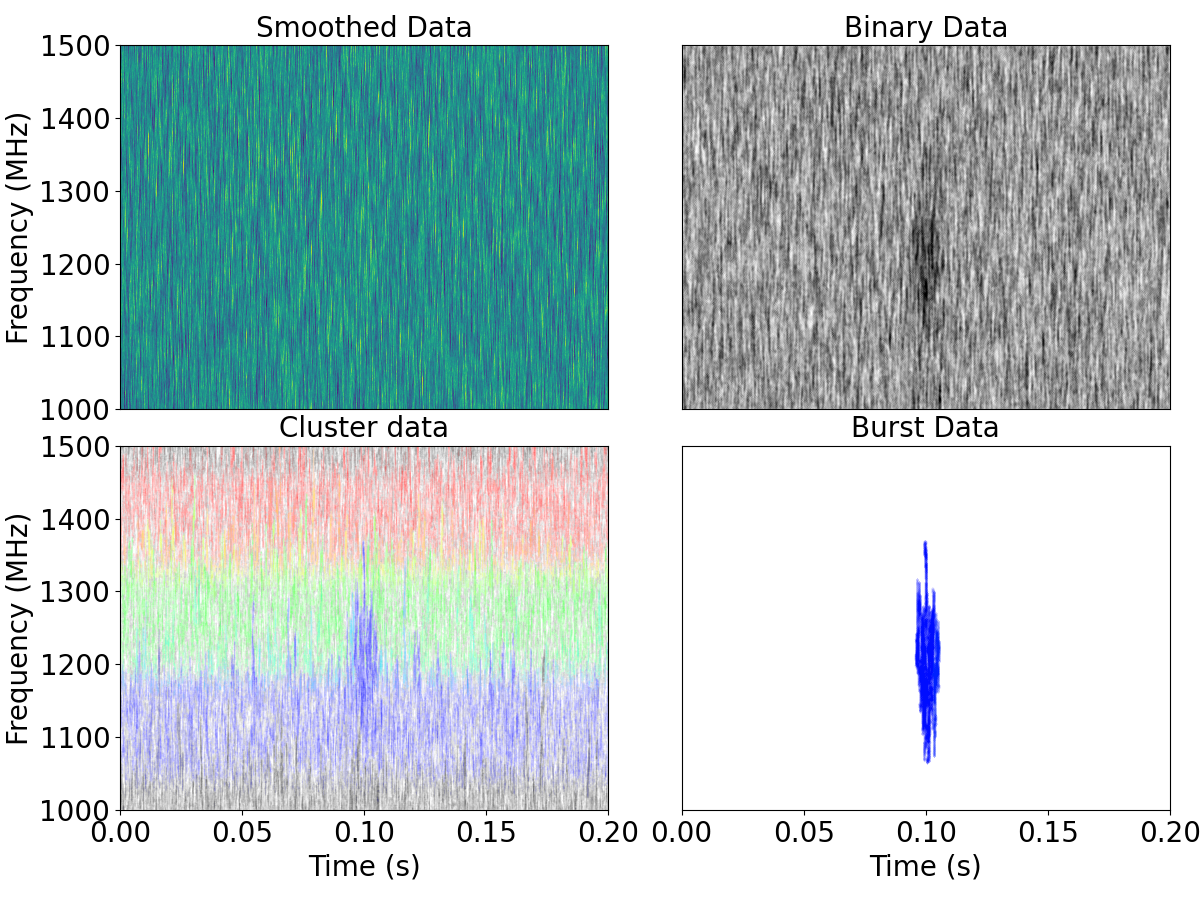}
\caption{A presentation of the detection result of DANCE to the data shown in Fig.\ref{simulate}. The progress DANCE first smooths the data and then filters the smoothed data according to the strength values of each pixel. The cluster data is derived by performing the DBSCAB on the binary data, and the final burst data is filtered by the size and width of the clusters. }
\label{simulate_cluster}
\end{figure}
Fig.\ref{simulate_cluster} is the detection result of the simulated signal above. The right bottom panel shows that DANCE finds this burst correctly. We use DANCE to process the simulated data, the total sample number ($SN$) is 2,550. Among these samples, the pulse width ranges from 1~ms to more than 10~ms; and the bandwidth ranges from 100~MHz to 500~MHz. The SNR varies from less than 2 to greater than 8. We counted the true positive samples and false positive samples and recorded the number of the two samples as $TN$ and $FN$. We use the concept of recall and precision to test the performance of DANCE on simulated FRBs. Because we do not set negative samples (i.e. a data set without simulating FRB signals), the recall and precision can be calculated as
\begin{align}
\begin{aligned} 
&Recall = \frac{TP}{SN};\\
&Precision = \frac{TP}{TP+FP}.
\end{aligned}
\end{align}

\begin{figure}
\includegraphics[width=0.5\columnwidth]{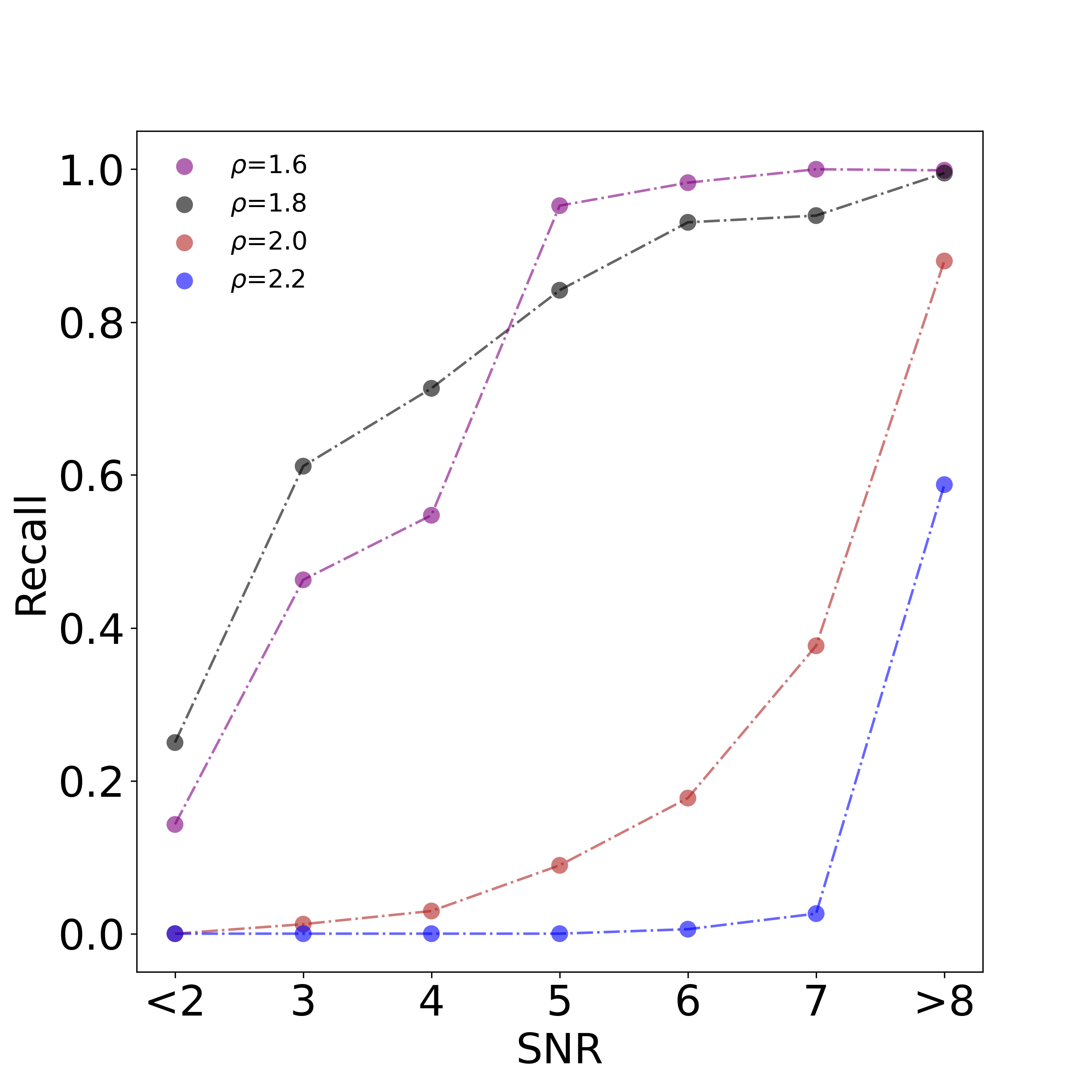}
\includegraphics[width=0.5\columnwidth]{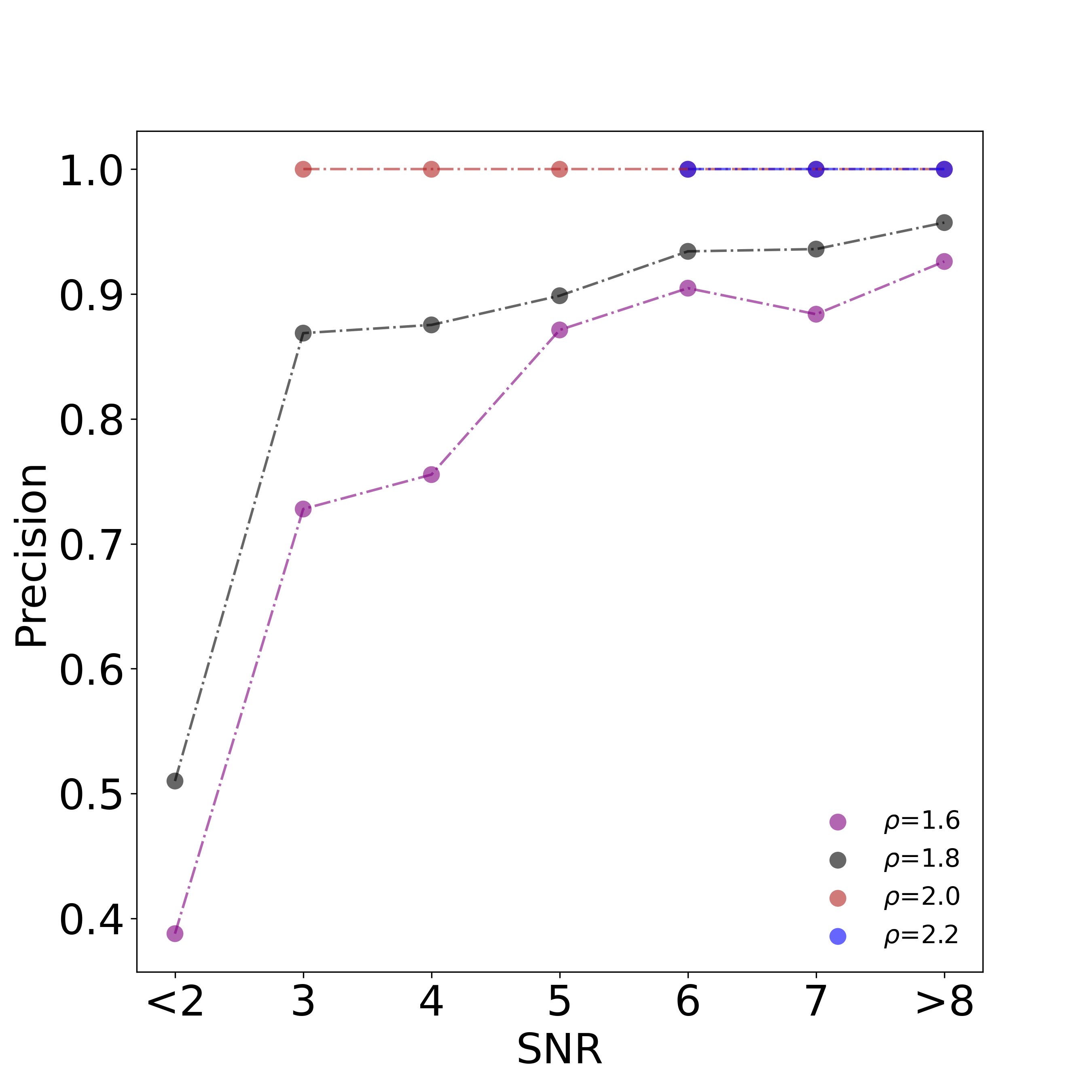}
\caption{Recall and precision of DANCE based on simulated FRBs.}
\label{perf}
\end{figure}

{We conducted a series of simulations to evaluate the performance of DANCE. The simulated bursts span an SNR range from below 2 to above 8. Two parameter sets were tested, with \textit{Eps} = 2 and \textit{Eps} = 4, respectively. For each \textit{Eps} value, we explored various density thresholds, $\rho$. Fig.\ref{perf} presents the recall and precision achieved with \textit{Eps} = 4 and $\rho = 1.6, 1.8, 2.0,$ and $2.2$. Since the detection sensitivity with \textit{Eps} = 4 is excessively high for bursts with SNR $\gtrsim 3$, we focus primarily on the results obtained with \textit{Eps} = 2. The results indicate that an appropriate density parameter should be smaller than 2. Considering bursts with SNR $\geq 5$ as representative of convincing FRB signals, the recall and precision reach 0.98/0.90 for $\rho = 1.6$ and 0.93/0.94 for $\rho = 1.8$.}

It should be noted that the recall and precision of signal detection based on simulation data are usually biased from real data, especially when the real data contains unstable RFI \citep{yuan2022categorize}. We will present the performance of DANCE on real FRBs in the next section.

\subsection{Detection on FRB20201124A}
We test the performance of DANCE with a real observation case of FRB~20201124A. This is a highly active repeater that was first discovered by the Canadian Hydrogen Intensity Mapping Experiment Fast Radio Burst (CHIME/FRB) \citep{2021ATel14497....1C}. The data we examined was observed by FAST on September 29th, 2021, at a period of heightened activity for this source. \citep{zhou2022fast,zhang2022fast,jiang2022fast,niu2022fast}. 

The observation time is 1 hour and the frequency band covers from 1000 to 1500~MHz. The time and frequency resolutions are 49.152 $\mu s$ and 0.122~MHz. We de-disperse at first and then split the data into subsets without any time or frequency down-sampling. Every subset remains the same dimension ($4096\times4096$) with the simulated data.
The parameters we adapt also remain the same with the experiment on simulated bursts.

\begin{figure*}
\includegraphics[width=1.5\columnwidth]{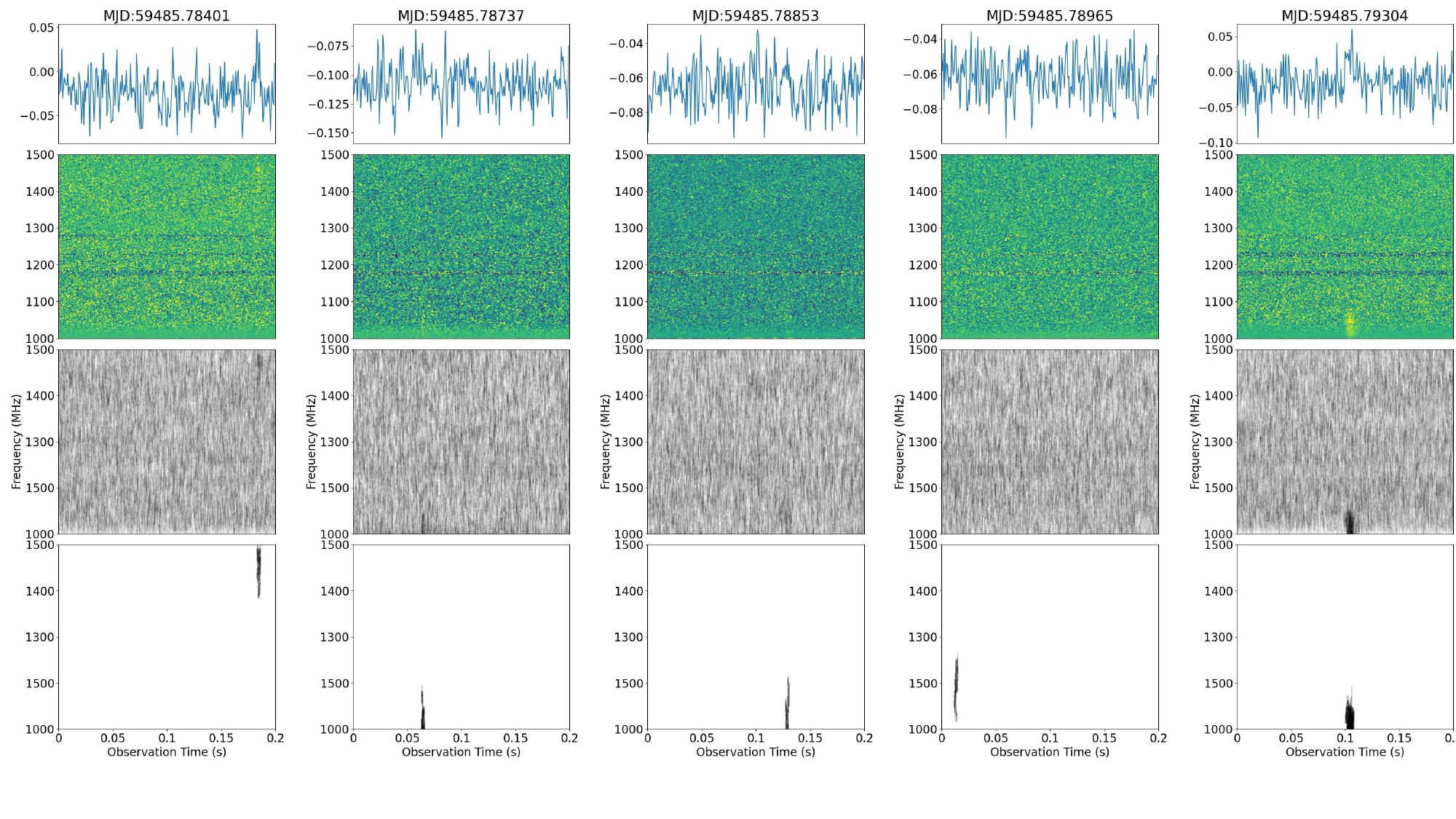}\\
\includegraphics[width=1.5\columnwidth]{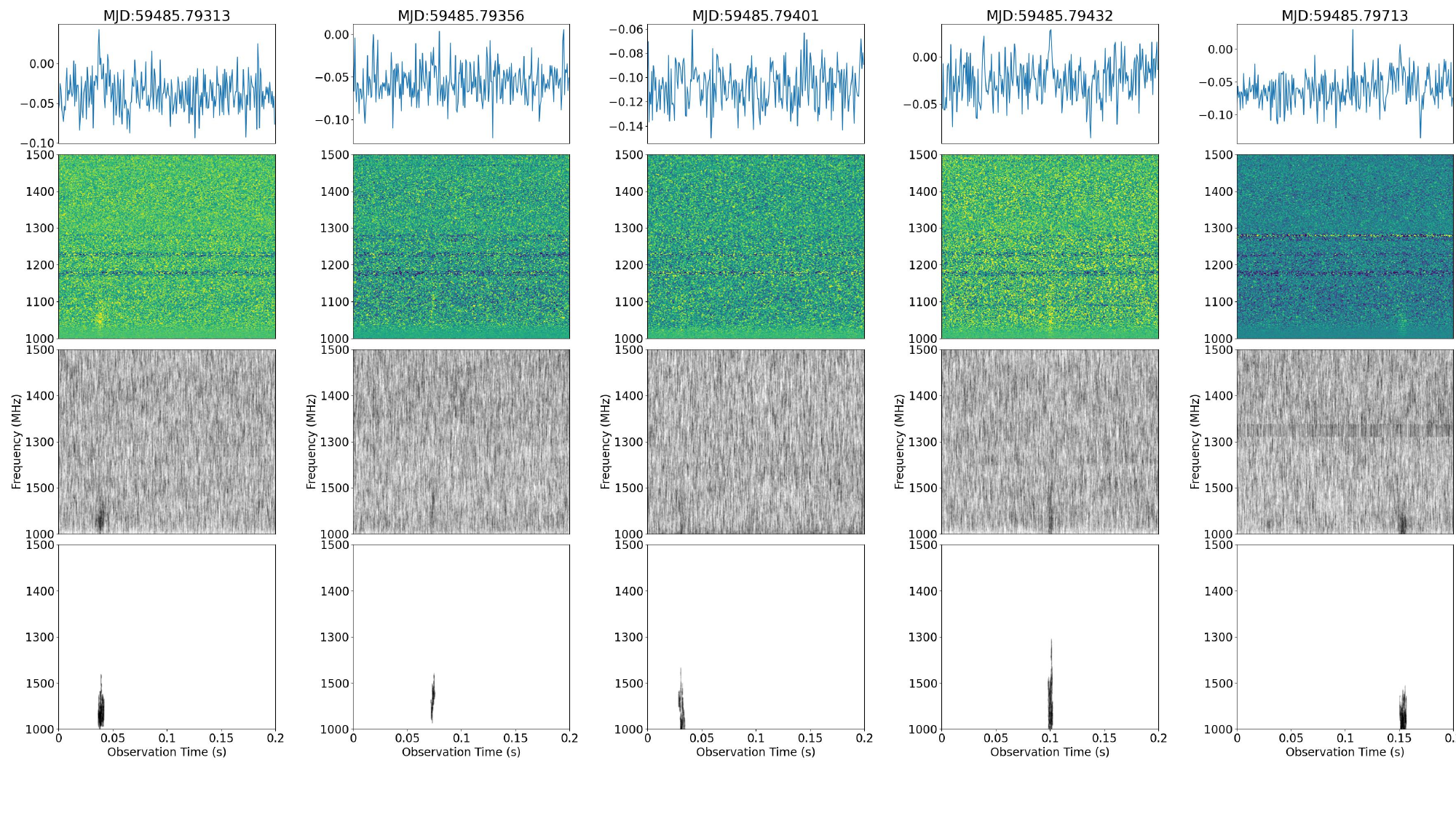}\\
\includegraphics[width=1.5\columnwidth]{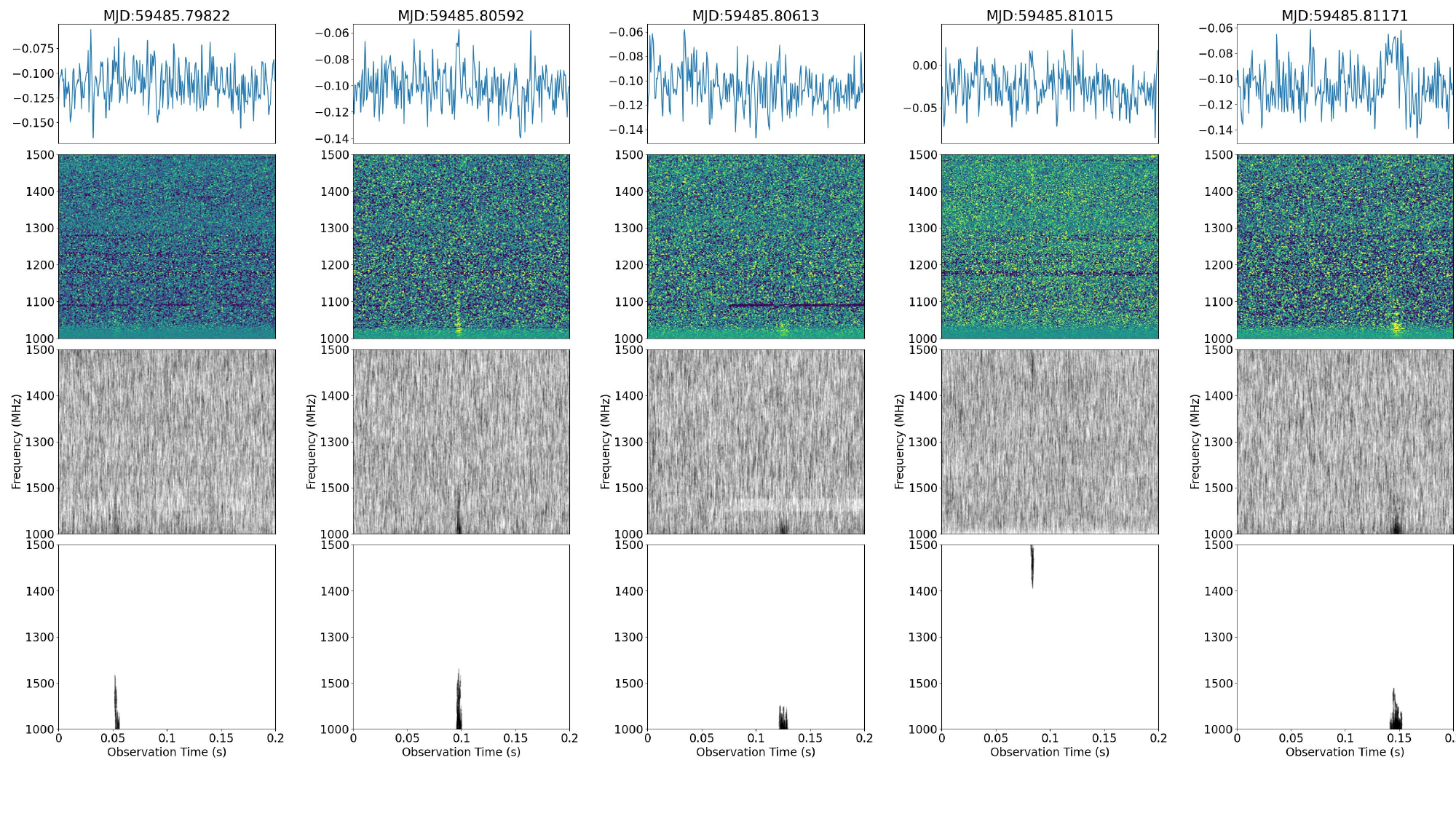}
\caption{True FRB signals --- isolated weak bursts. In all of these subfigures, the top panels are integrated profiles of the RFI-cleaned data, which have been 16-downsampled in both the frequency and time domains to provide a better representation of the bursts shown below the profiles. The bottom two panels show the signal density representations and the extracted burst labels. Because these bursts appear in isolation within 0.2 s, we named these bursts as isolated weak bursts.}
\label{burst1}
\end{figure*}

\begin{figure*}
\includegraphics[width=1.5\columnwidth]{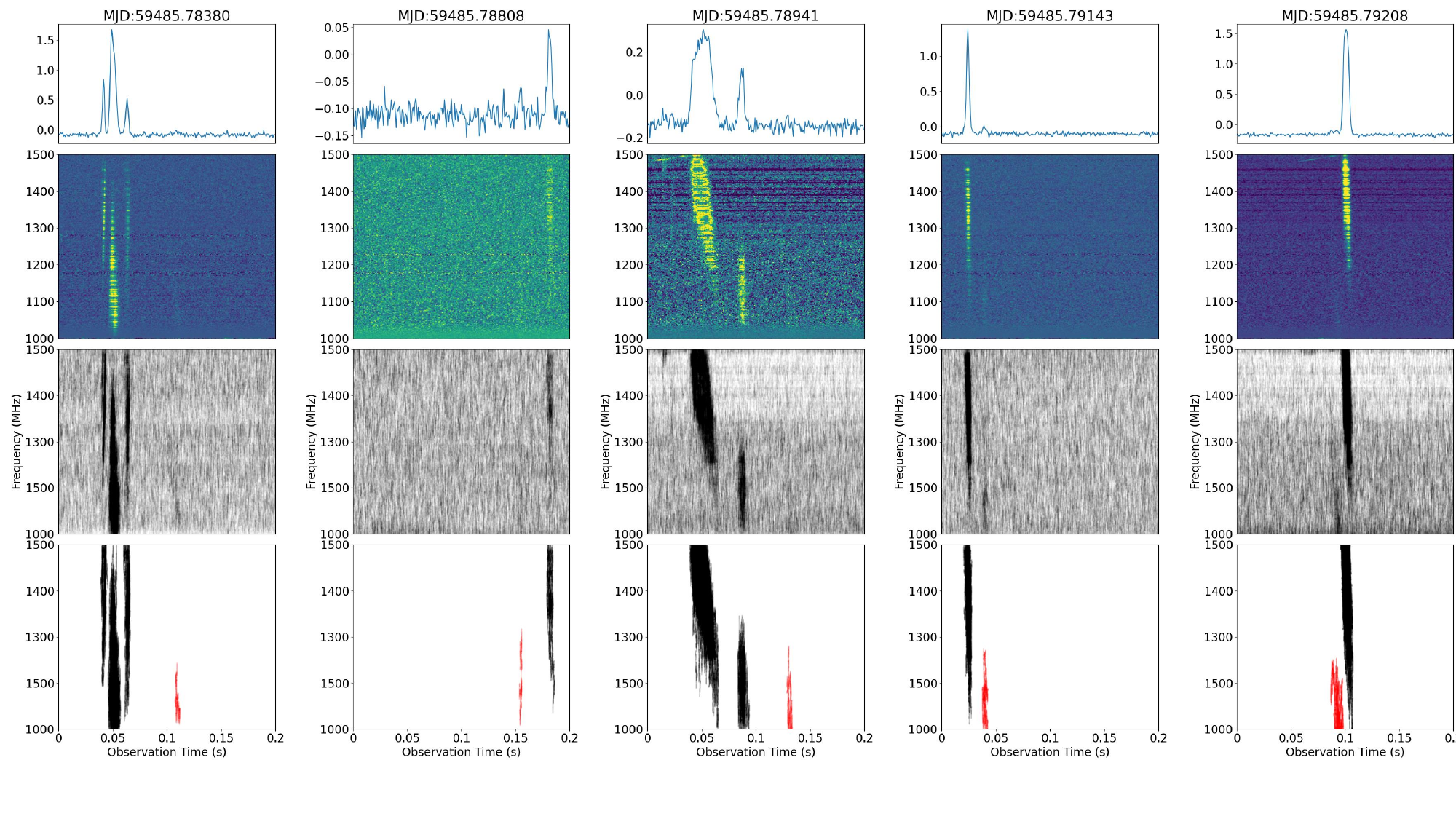}\\
\includegraphics[width=1.5\columnwidth]{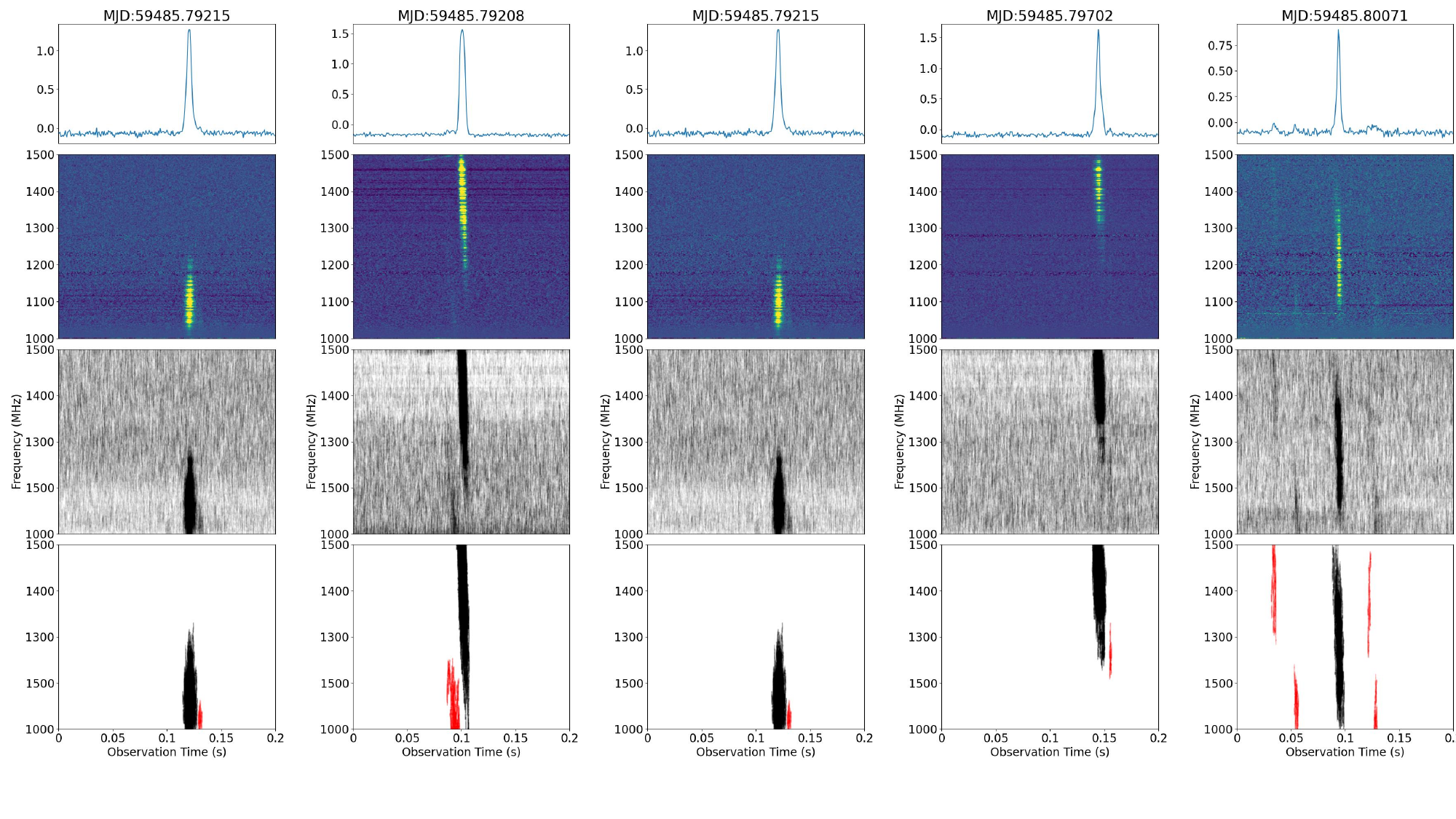}\\
\includegraphics[width=1.5\columnwidth]{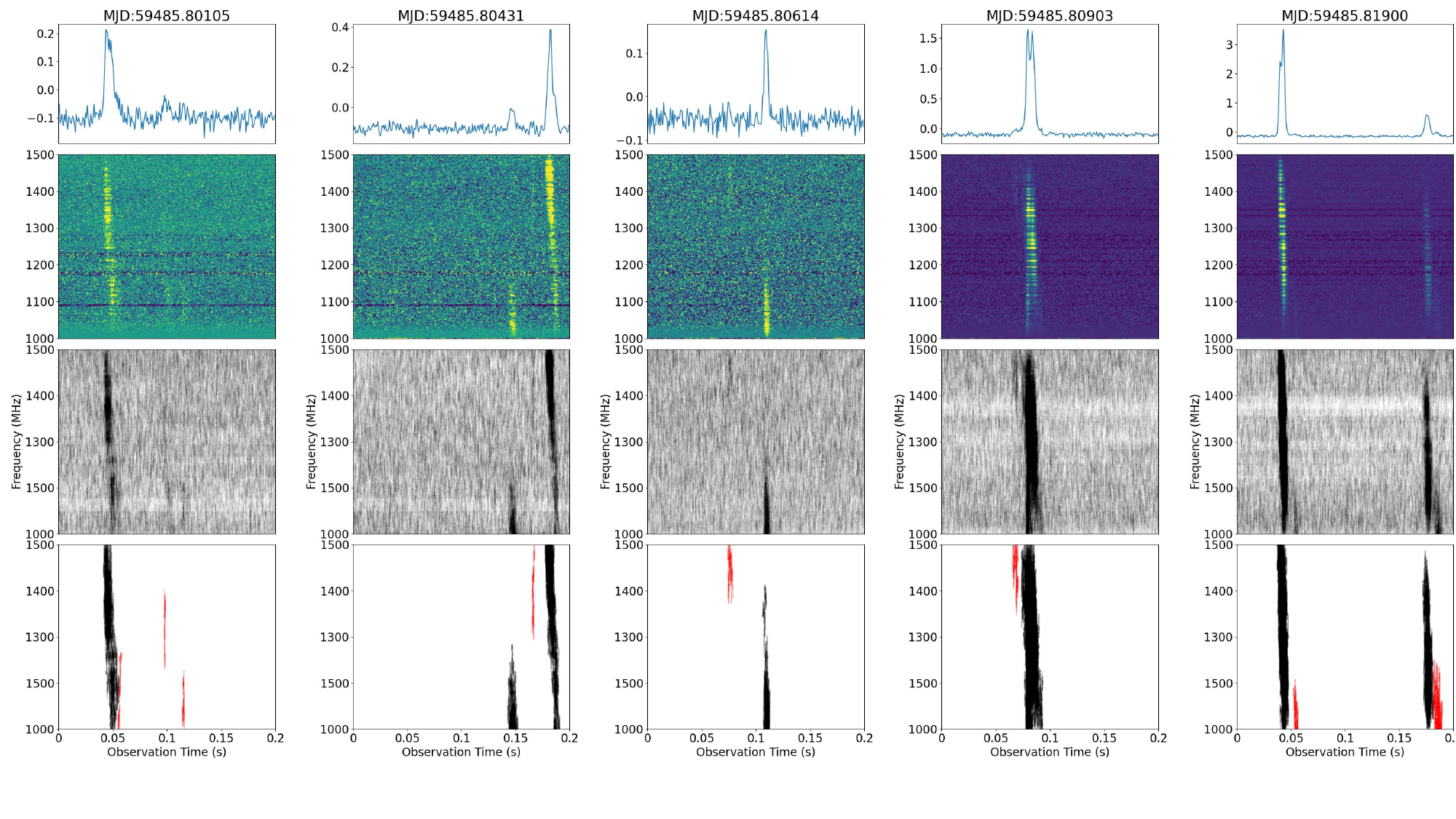}
\caption{True FRB signals --- associated weak bursts. The difference between these panels and the figures in Fig.\ref{burst1} is that weak bursts here (red clusters) are accompanied by relatively strong bursts (black clusters) within a character timescale like 0.2 s. We defined such bursts as associated weak bursts.}
\label{burst2}
\end{figure*}

We totally processed 17,263 data sets (0.2~s for each one), and the positive samples are 4000. The parameters for cluster forming and extraction are $Eps=2$; $MinPts=6$ and $Z=3$. We down-sampled every positive data set to a $256\times 256$ spectrum and then identified the real burst by eye. We confirmed 652 true bursts from the positive triggers at last. As a comparison, the number of reported bursts in this data set is 500 \citep{zhang2022fast}.
Among the signals we certified, there are many extremely weak signals whose SNR is far below 3$\sigma$. Fig.\ref{burst1} and \ref{burst2} show parts of the weak bursts. Interestingly, these weak bursts are not only isolated within a short timescale (0.2~s) but also closely leading or following strong bursts. We defined these two types of weak bursts as isolated and associated weak bursts. Such pre- or post-cursor weak burst components are also mentioned in other repeaters, like FRB 2020121102 \citep{caleb2020simultaneous}.

The successful capture of weak FRBs at the noise level is the advantage of DANCE over approaches based on integrated profile strength. Nevertheless, we also notice that the precision is not comparable with the experiments on simulated data. Misidentifications could be an unavoidable cost for weak FRB extraction as the RFI in real data is more complex than simulated Gaussian noise. Indeed, a high parameter level will definitely improve the precision, but it will also lead to an under-reporting of parts of weak FRBs. We suggest that it is worthwhile to sacrifice precision to improve the detection ability of weak signals, especially since the DANCE provides a visual advantage that greatly reduces the manual confirmation work.

\section{Discussion}
\label{sec:dis}
DANCE shows high sensitivity in identifying weak FRBs that remain indiscernible in conventional profiles. However, these target signals are de-dispersed, and the detection performance depends critically on the choice of parameters. In this section, we discuss the trials of blind searching of FRBs under different parameter settings.

The two key parameters, \textit{Eps} and $\rho$ (i.e. the \textit{MinPts}), determine the minimum spatial extent and density threshold required for a cluster to be identified. For a randomly distributed SDS, assuming that the point density is $\rho 0$ per pixel ($\rho 0 =0.25$ when the spectrum is filtered at a 1.5 IQR level), and given clustering parameters \textit{Eps} and $\rho$, then the {probability} of forming a cluster in the area is 
\begin{align}
    P_{S,\rho} = C_{ \lceil S  \rceil}^{ \lceil S\rho \rho 0  \rceil} \rho 0 ^{\lceil S\rho \rho 0  \rceil},
\end{align}
where $S\sim \pi {Eps}^2$. Theoretically, increasing either \textit{Eps} or $\rho$ reduces the probability of noise-induced clustering, thereby improving the reliability of true signal identification. Meanwhile, overly large parameter values may also suppress weak or small-scale signals.

\begin{figure}
\includegraphics[width=0.98\columnwidth]{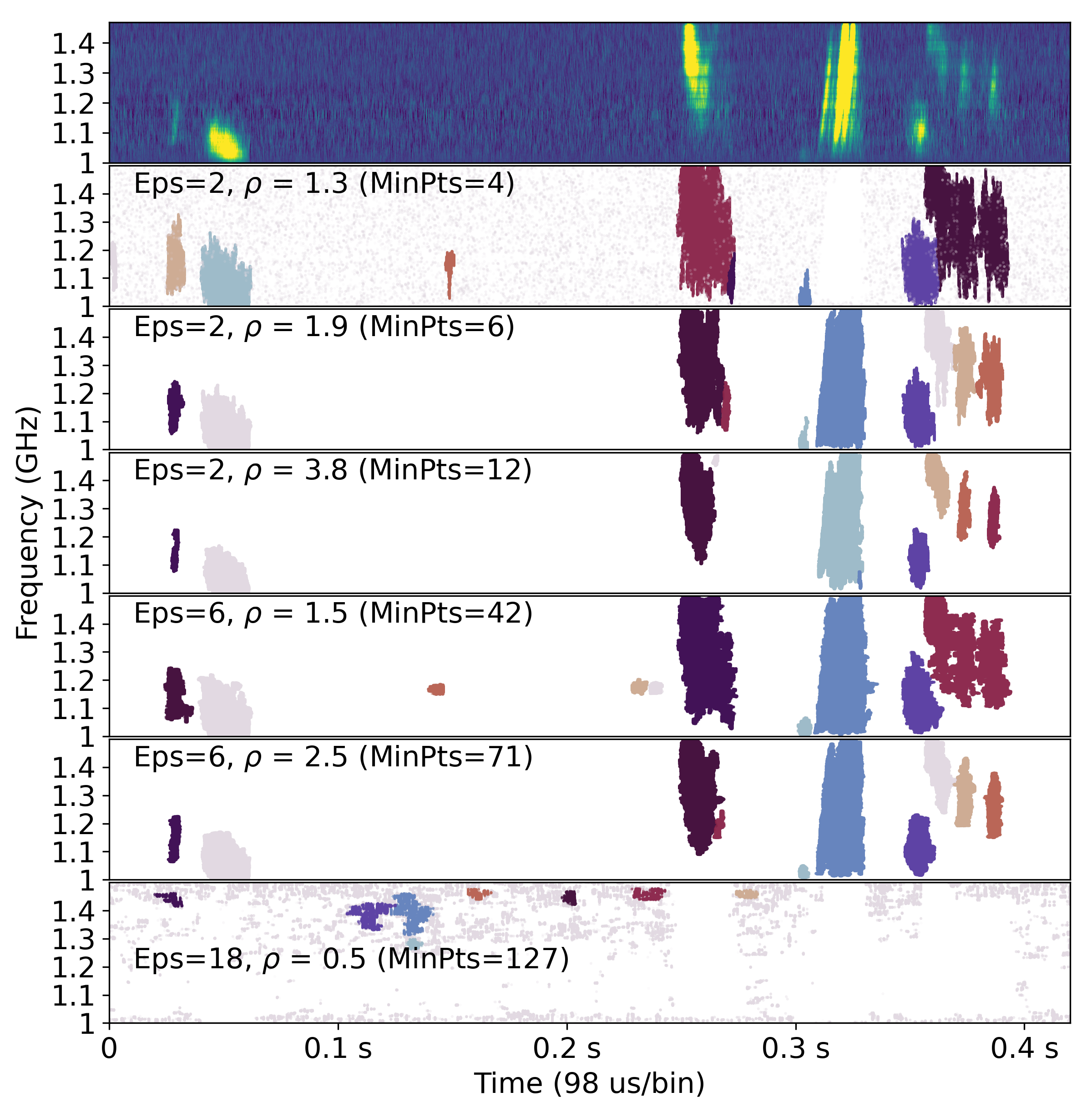}
\caption{{Clustering of FRBs under different parameter settings. The top panel shows a smoothed spectrum containing several bursts, while the six bottom panels display the ten most significant clusters identified under varying (\textit{Eps}, $\rho$) combinations, as indicated in each panel.}}
\label{add}
\end{figure}

{In practice, adopting multiple parameter combinations can be advantageous when dealing with complex burst morphologies. Fig.\ref{add} illustrates clustering results for several bursts with distinct bandwidths, durations, and intensities. A suggested detection approach, using \textit{Eps}~$=2,\rho=1.9$ successfully identifies all isolated structures, while a second configuration with \textit{Eps}~$=6$ and $\rho=2.5$ detects all bursts as well, and producing clusters confined to more precise time–frequency regions.}

\begin{figure*}
\includegraphics[width=2\columnwidth]{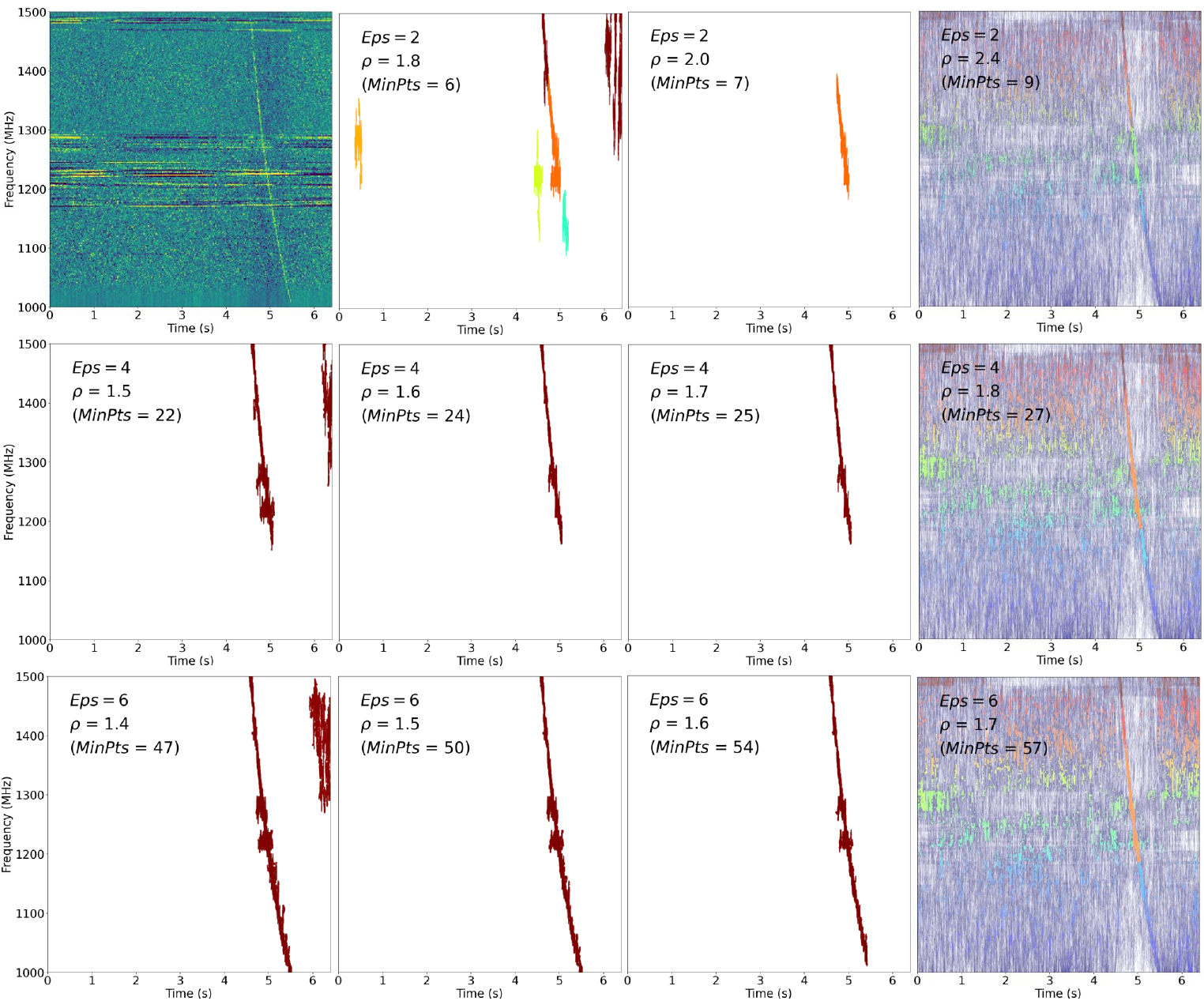}
\caption{DANCE performance on non-dispersed FRB signals. The top-left sub-panel displays the original spectrum. The three sub-panels in the right column present all identified clusters of the spectrum using different clustering parameters before final extraction. The remaining sub-panels present the filtered clusters as the final results.}
\label{non-disp}
\end{figure*}

We use DANCE to extract a non-dispersed FRB under different cluster parameters, shown in Fig.\ref{non-disp}. The left top panel shows the FRB signal and the other panels are clusters extracted by DANCE. Among the cluster panels, the three in the right column are the whole clusters of DBSCAN. We present all the clusters because DANCE detected no positive samples under the corresponding parameters. 

This burst is the sample shown in the top left panel of Fig.\ref{burst2} (MJD=59485.78380). We downsample the spectrum to a 4096$\times$4096 matrix, the time resolution is 0.2~s. Thus the original triple components burst was compressed together. The \textit{Eps} and \textit{MinPts} set for DBSCAN were labeled in each panel. The cluster extraction threshold $Z$ is 3. For each group of \textit{Eps}, low density causes sporadic noise or RFI clusters (left panels), and high density leads to discrete FRB clusters, which have small sizes and can not pass the extraction threshold.

The clustering to non-dispersed FRB demonstrates the {capability} of DANCE for FRB searches. Using this method, one can detect a pulse without {performing} de-dispersion. That is, we can use DANCE search FRB or other transit pulses without much computational resources and {providing immediate visual assessment}, {and provide a rapid initial check for the presence of transient bursts in an observation}. Nevertheless, there are limitations of DANCE in bland searches. {First, it is parameter-dependent: FRBs with different DMs have varying distributions in the time-frequency domain, so DANCE may be triggered under different density-reachability and connectivity conditions.} Second, only strong pulses might be clustered in relatively larger scale and density parameters. Third, DANCE does not provide DM information.  

{This limitation arises from the intrinsic behavior of the underlying DBSCAN algorithm, whose effectiveness strongly depends on the geometric morphology of the target signal\citep{ester1996density}. DBSCAN defines clusters as regions of high point density in a metric space, i.e. points shared a same cluster are neighbored in Euclidean distance, which defines as $dist_{\mathrm{p,q}} = \sqrt{(x_p-x_q)^{2}+(y_p-y_q)^{2}}$. Thus a neighbouring clustered samples approximately forms an isotropic circular or elliptical zone in the feature space. In the case of non-dedispersed signals, the intrinsic geometry of the burst is highly anisotropic, appearing as a long, tilted track across the time–frequency plane due to dispersion. This mismatch between the isotropic distance metric and the anisotropic signal structure can lead to cluster fragmentation or reduced recall, as distant parts of the same burst fall outside the defined neighborhood radius $Eps$.}

{Furthermore, in non-dedispersed data, the effective density of signal points projected along the dispersion track is substantially lower than that in dedispersed data. Consequently, achieving the $MinPts$ threshold for core point identification becomes more difficult. Increasing the neighborhood size $Eps$ can partially compensate for this effect, but it also increases the inclusion of noise, thereby reducing precision. Consequently, DBSCAN---and thus DANCE--has limited sensitivity to weak, dispersed bursts; typically, only strong initial pulses can be clustered with a positive response of DANCE. Moreover, since different dispersion slopes correspond to distinct track geometries, the detection sensitivity may vary even for bursts of comparable intrinsic strength. It prevents a unified quantitative assessment of DANCE’s performance on non-dedispersed data.}

{Despite these limitations, DANCE remains a useful tool for preliminary detection in raw, non-dedispersed data. It can efficiently highlight strong transient candidates without the need for computationally expensive dedispersion trials. Once such candidates are triggered, a subsequent dedispersed, post-processed search can be conducted with optimized parameters to achieve higher sensitivity and completeness. This two-step strategy effectively combines DANCE’s robustness with improved performance after dispersion correction.
}
\section{conclusion}
\label{sec:con}
We developed an unsupervised method, DANCE, to detect and isolate FRB signals in the original spectrum, {providing} visual identification of their precise time-frequency locations and detailed morphology. DANCE demonstrates reliable accuracy in detecting weak FRBs; simulations indicate that for FRBs with $SNR > 5$, the detection precision reaches 93\%. The practical performance on real observation shows that DANCE has {uncovered} numerous previously undetected weak, narrow-band FRBs in historical datasets. This method shows promise as a {powerful} tool for identifying and distinguishing weak FRBs, even in cases where they are located very close to stronger signals.

\section*{Acknowledgements}

This work made use of data from the FAST, a Chinese national
mega-science facility, built and operated by the National
Astronomical Observatories, Chinese Academy of Sciences. This work is supported by the independently funded research program of National Space Science Center (``2025 Young Scientist Innovation Program'', No. E5PD40004S).
This work is supported by the National Natural Science Foundation of China (NSFC, No. 12503055) and the Postdoctoral Fellowship Program of CPSF under Grant Number GZB20250737. This work is also supported by the National SKA Program of China (Nos. 2020SKA0120200 and 2020SKA0120100).

\section*{Data Availability}
The data supporting this article will be shared when a reasonable request is made to the corresponding author. The raw data and can be accessed from FAST data center \url{https://fast.bao.ac.cn/}. The code used in this article is publicly available at: 
\url{https://github.com/Yuanao/DANCE_v-test}.

\newpage
\bibliographystyle{mnras}
\bibliography{ref} 






\bsp	
\label{lastpage}
\end{document}